\documentclass[twocolumn,pra,10pt]{revtex4}
\usepackage{graphicx,amsmath,amssymb,xspace,ntheorem,hyperref}
\usepackage{color}
\usepackage[T1]{fontenc}

\begin{document}

\newcommand{\ketbra}[2]{| #1\rangle \langle #2|}
\newcommand{\ket}[1]{| #1 \rangle}
\newcommand{\bra}[1]{\langle #1 |}
\newcommand{\Tr}{\mathrm{Tr}}
\newcommand\F{\mbox{\bf F}}
\newcommand{\h}{\mathcal{H}}
\newcommand{\braket}[2]{\langle #1|#2 \rangle}

\newcommand{\PSD}{\textup{PSD}}

\newcommand{\C}{\mathbb{C}}
\newcommand{\X}{\mathcal{X}}
\newcommand{\Y}{\mathcal{Y}}
\newcommand{\Z}{\mathcal{Z}}
\newcommand{\sspan}{\mathrm{span}}
\newcommand{\kb}[1]{\ket{#1} \bra{#1}}
\newcommand{\pos}{D}

\newcommand{\thmref}[1]{\hyperref[#1]{{Theorem~\ref*{#1}}}}
\newcommand{\lemref}[1]{\hyperref[#1]{{Lemma~\ref*{#1}}}}
\newcommand{\corref}[1]{\hyperref[#1]{{Corollary~\ref*{#1}}}}
\newcommand{\eqnref}[1]{\hyperref[#1]{{Equation~(\ref*{#1})}}}
\newcommand{\claimref}[1]{\hyperref[#1]{{Claim~\ref*{#1}}}}
\newcommand{\remarkref}[1]{\hyperref[#1]{{Remark~\ref*{#1}}}}
\newcommand{\propref}[1]{\hyperref[#1]{{Proposition~\ref*{#1}}}}
\newcommand{\factref}[1]{\hyperref[#1]{{Fact~\ref*{#1}}}}
\newcommand{\defref}[1]{\hyperref[#1]{{Definition~\ref*{#1}}}}
\newcommand{\exampleref}[1]{\hyperref[#1]{{Example~\ref*{#1}}}}
\newcommand{\hypref}[1]{\hyperref[#1]{{Hypothesis~\ref*{#1}}}}
\newcommand{\secref}[1]{\hyperref[#1]{{Section~\ref*{#1}}}}
\newcommand{\chapref}[1]{\hyperref[#1]{{Chapter~\ref*{#1}}}}
\newcommand{\apref}[1]{\hyperref[#1]{{Appendix~\ref*{#1}}}}
\newcommand\rank{\mbox{\tt {rank}}\xspace}
\newcommand\prank{\mbox{\tt {rank}$_{\tt psd}$}\xspace}
\newcommand\alice{\mbox{\sf Alice}\xspace}
\newcommand\bob{\mbox{\sf Bob}\xspace}
\newcommand\pr{\mbox{\bf Pr}}
\newcommand\av{\mbox{\bf{\bf E}}}
\newcommand{\pabxy}{(p(ab|xy))}
\newcommand{\calQ}{\mathcal{Q}}
\def\be{\begin{equation}}
\def\ee{\end{equation}}

\newcommand{\comment}[1]{{}}
\newcommand{\blue}[1]{\textcolor{blue}{#1}}
\newcommand{\red}[1]{\textcolor{red}{#1}}

\title{\vspace{-1cm} Experimental Entanglement Quantification for Unknown Quantum States in a Semi-Device-Independent Manner}

\author{{Yu Guo}$^{1,2}$}
\author{{Lijinzhi Lin}$^{3}$}
\author{{Huan Cao}$^{1,2}$}
\author{{Chao Zhang}$^{1,2}$}
\author{{Xiaodie Lin}$^{3}$}
\author{{Xiao-Min Hu}$^{1,2}$}
\author{{Bi-Heng Liu}$^{1,2,}$}\email{bhliu@ustc.edu.cn}
\author{{Yun-Feng Huang}$^{1,2}$}
\author{{Zhaohui Wei}$^{3,}$}\email{weizhaohui@gmail.com}
\author{{Yong-Jian Han}$^{1,2,}$}\email{smhan@ustc.edu.cn}
\author{{Chuan-Feng Li}$^{1,2,}$}\email{cfli@ustc.edu.cn}
\author{{Guang-Can Guo}$^{1,2}$}

\affiliation{$^{1}$CAS Key Laboratory of Quantum Information, University of Science and Technology of China, Hefei, 230026, People's Republic of China\\$^{2}$CAS Center For Excellence in Quantum Information and Quantum Physics, University of Science and Technology of China, Hefei, 230026, People's Republic of China\\$^{3}$Center for Quantum Information, Institute for Interdisciplinary Information Sciences, Tsinghua University, Beijing 100084, People's Republic of China}

\begin{abstract}
Using the concept of non-degenerate Bell inequality, we show that quantum entanglement, the critical resource for various quantum information processing tasks, can be quantified for any unknown quantum states in a semi-device-independent manner, where the quantification is based on the experimentally obtained probability distribution and beforehand knowledge on quantum dimension only. Specifically, as an application of our approach on multi-level systems, we experimentally quantify the entanglement of formation and the entanglement of distillation for qutrit-qutrit quantum systems. In addition, to demonstrate our approach for multi-partite systems, we further quantify the geometry measure of entanglement of three-qubit quantum systems. Our results supply a general way to reliably quantify entanglement in multi-level and multi-partite systems, thus paving the way to characterize many-body quantum systems by quantifying involved entanglement.
\end{abstract}

\maketitle

{\em Introduction.---}Quantum entanglement, the key resource for quantum communication~\cite{Gisin07} and quantum key distribution~\cite{BB84,Ekert92,Gisin02}, provides remarkable quantum advantage for quantum simulators and quantum computers over their classical counterparts~\cite{Vidal03,oneway}. It also supplies critical information about many-body physics, such as the thermalization~\cite{Srednicki94, rmp19}, the many-body localization~\cite{bloch15,rmp19}, and topological order~\cite{xiaogang95,xiaogang06,kitaev06}. Therefore, efficiently quantifying entanglement is one of the main tasks in quantum information and many-body quantum physics.

Quantum entanglement witness~\cite{Guhne02} is widely used to detect genuine entanglement of quantum systems. However, it is unsatisfactory for the following reasons: firstly, certain accurate information about the target state is needed \cite{Brunner08}, which prevents its application to unknown states; secondly, from experimental aspect, the exact knowledge on the measurement device needed by the approach is impossible to obtain; lastly, but not least, quantum entanglement witnesses usually only detect the presence of entanglement, which is insufficient for many applications such as classifying the topological phases in many-body systems by entanglement~\cite{xiaogang06,kitaev06}.

The device-independent (DI) method, initially introduced in quantum key distribution~\cite{Acin07} and self-testing~\cite{Mayers04}, can also be used to detect the entanglement of a state, where the detection is based only on the corresponding Bell-type correlations generated by locally measuring the target state experimentally~\cite{Moroder13}, and all the involved devices are regarded as black boxes (i.e. we do not have to care about internal workings of the quantum devices). As a result, this approach can overcome the critical drawbacks of the entanglement witness method mentioned above. In fact, the DI method has been experimentally implemented to demonstrate dimension witness~\cite{DIDW1,DIDW2}, Bell-inequality violation~\cite{LHFBN}, randomness generation~\cite{DIRG}, and self-testing~\cite{Zhang18}. Furthermore, measurement-DI~\cite{LCQ12,BP12} and semi-DI schemes~\cite{MG12,LVB11}, where partial information on the target system is known reliably, have also been extensively studied. For example, semi-DI schemes assume that quantum dimension is known reliably before characterizing the target unknown quantum system.

In this paper, we experimentally demonstrate that the semi-DI method can be utilized to efficiently quantify entanglement in multi-level and many-body quantum systems. Particularly, since the foundation of our method is Bell-type correlations, whose size is not determined by quantum dimension directly, the number of quantum measurements needed is very modest, implying that our method is very efficient.

More specifically, with the help of the Collins-Gisin-Linden-Masser-Popescu (CGLMP) inequality~\cite{CGP+02}, we quantify \emph{the entanglement of formation} and \emph{the entanglement of distillation} in qutrit-qutrit systems based only on the experimentally obtained probability distribution, demonstrating our approach on multi-level systems. In addition, as a demonstration of multi-partite entanglement quantification, we further quantify \emph{the geometric measure of entanglement} in 3-qubit systems by examining experimentally obtained probability distributions with the Mermin-Ardehali-Belinskii-Klyshko (MABK) inequality~\cite{Mermin90,Ardehali92,BK93}. We would like to stress that our method is general for multi-level and many-body systems, thus paves the way to study many-body physics through efficiently quantifying its entanglement.

{\em Overview of the theory.---}Suppose $\rho$ is an $n$-partite quantum state, for each set of local measurements $\vec{x}\equiv (x_1,x_2,...,x_n)$ ($x_i\in X_i,i=1,2,...,n$ and $X_i$ is the set of von Neumann measurements on the $i$-th party) measured on each partite, their outcomes are denoted as $\vec{a}\equiv (a_1,a_2,...,a_n)$ ($a_i\in A_i$ and $A_i$ is the set of the possible outcomes of the measurement $x_i$). These local measurements generate a quantum correlation expressed as the probability distribution $p(\vec{a}|\vec{x})=\Tr((\bigotimes\limits_{i=1}^{n}M_{x_i}^{a_i})\rho)$ where $M_{x_i}^{a_i}$ is the measurement operator with outcome $a_i$ for the measurement $x_i$ performed on the $i$-th party. For convenience, we denote the combination of these local measurements as $\{M_{\vec{x}}\}$. The probability distribution $p(\vec{a}|\vec{x})$ can be directly obtained in experiment and can be used to detect nonlocality. Here, we further use them to quantify the entanglement of unknown quantum states of known dimension, that is, in a semi-DI fashion.

To quantify entanglement of a multi-partite quantum state, a general measure is needed and we choose the geometric measure of entanglement (GME) ~\cite{BH01,WG03}. The GME of a general quantum state $\rho$ is defined by convex roof construction as:
\begin{equation*}
E_G(\rho)\equiv 1-\max_{\rho=\sum_ip_i\ketbra{\psi_i}{\psi_i}}\sum_{i}p_i\sup_{\ket{\phi_i}\in\text{sep}_n}|\braket{\psi_i}{\phi_i}|^2,
\end{equation*}
where $\text{sep}_n$ is the set of $n$-partite product pure states.

To obtain the GME from $p(\vec{a}|\vec{x})$, we need information about two fundamental quantities: the maximal overlap between $\rho$ and a pure product state $|\phi\rangle$, and the purity of $\rho$ (it means how it close to a pure state, defined as $\Tr(\rho^2)$). Fortunately, an upper bound for the former, denoted as $\hat{F}$, can be directly fulfilled by numerical approaches like the shifted higher-order power method (SHOPM) algorithm~\cite{KM11} from the distribution $p(\vec{a}|\vec{x})$ ~\cite{LW20} (see Appendix B for more details). Meanwhile, a lower bound for the purity of $\rho$ can also be obtained directly from the distribution $p(\vec{a}|\vec{x})$, if one applies the concept of non-degenerate Bell inequalities~\cite{WL19}.

After choosing a set of local measurement ($\{M_{\vec{x}}\}$) and the corresponding outcomes ($\{\vec{a}\}$), a general Bell inequality can be expressed as $I(\rho,\{M_{\vec{x}}\},\{\vec{a}\})=\sum_{\vec{a},\vec{x}}c_{\vec{x}}^{\vec{a}}p(\vec{a}|\vec{x})\leq C_l$, where $c_{\vec{x}}^{\vec{a}}$ are real numbers and $C_l$ is the maximal classical value. Intuitively, if a quantum state $\rho$ remarkably violates the Bell inequality $I(\rho,\{M_{\vec{x}}\},\{\vec{a}\})\leq C_l$, we hope $\rho$ can be certified to be close to a pure state, i.e., the purity $\Tr(\rho^2)$ is close to 1, like in the Clauser-Horne-Shimony-Holt (CHSH) inequality ~\cite{CHSH70}. The concept of non-degenerate for Bell inequalities is used to make this intuition strict. Explicitly, suppose the target quantum system has a dimension vector $\vec{d}\equiv (d_1,d_2,...,d_n)$ (i.e. $d_i$ is the dimension of the $i$-th party), $I(\rho,\{M_{\vec{x}}\},\{\vec{a}\})\leq C_l$ is called non-degenerate, if there exist two real numbers $0\leq\epsilon_1<\epsilon_2\leq C_q(\vec{d})$ ($C_q(\vec{d})$ is the maximal value of the Bell expression for quantum systems of given dimension vector $\vec{d}$) such that, for any two orthogonal quantum states $\ket{\alpha}$ and $\ket{\beta}$, $I(\ketbra{\alpha}{\alpha},\{M_{\vec{x}}\},\{\vec{a}\}) \geq C_q(\vec{d})-\epsilon_1$ always implies that
$I(\ketbra{\beta}{\beta},\{M_{\vec{x}}\},\{\vec{a}\})\leq C_q(\vec{d})-\epsilon_2$. In fact, many notable Bell inequalities, such as the MABK inequality in qubit systems and the CGLMP inequality in qutrit systems, have been proved to be non-degenerate~\cite{SG01,LW20,WL19}.

\begin{figure*}[bhtp]
\begin{center}
\includegraphics [width= 1.4\columnwidth]{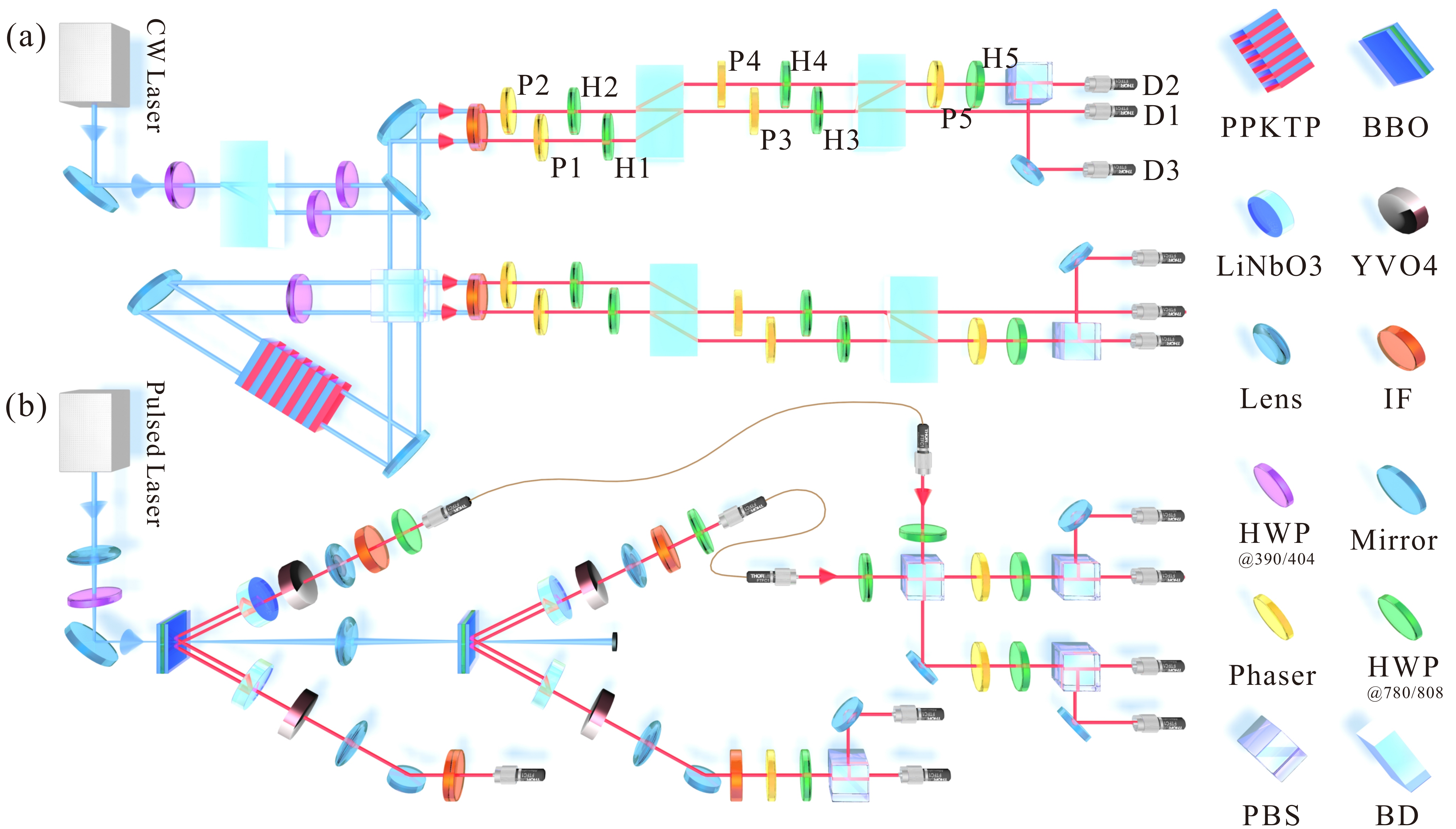}
\end{center}
\caption{Experimental setup for the semi-DI entanglement quantification of (a) qutrit-qutrit and (b) three-qubit entangled states, both of which can be decomposed into an entangled source and a measurement apparatus. (a) An entangled photon pair is generated from SPDC at a type-II cut periodically poled KTP (PPKTP) crystal embedded in a two-path Sagnac interferometer and pumped by a continuous-wave violet laser (power is 4~mW, working at 404~nm). Qutrit-qutrit states are encoded in the hybrid of the path and polarization degrees of freedom of the photons. The measurement settings for the CGLMP inequality can be implemented via the configuration composed of a series of Phasers (combination of two QWPs and an HWP), HWPs, BDs, and PBS. (b) Polarization encoded three-photon GHZ states are produced by combining two pairs of entangled photons generated from two sandwichlike BBO crystals pumped by a ultraviolet laser (with a central wavelength of 390~nm, a pulse repetition rate of 80~MHz and a power of 25~mW). LiNbO3 and YVO4 are used for spatial and temporal compensations between horizontal and vertical polarizations respectively. IF: interference filter; HWP: half-wave plate; QWP: quarter-wave plate; PBS: polarizing beam splitter; BD: beam displacer.
}
\label{fig:1}
\end{figure*}

Suppose the Bell inequality $I(\rho,\{M_{\vec{x}}\},\{\vec{a}\})\leq C_l$ is non-degenerate with parameters $\epsilon_1$ and $\epsilon_2$,  $\rho$ has an orthogonal decomposition $\rho=\sum_{i}a_i\ket{\psi_i}\bra{\psi_i}$, and $I(\rho,\{M_{\vec{x}}\},\{\vec{a}\})\geq C_q(\vec{d})-\epsilon_1$, then, according to the definition of the non-degenerate, it can be proved that there exists $a_i$ such that $a_i\geq 1-\epsilon_1/\epsilon_2$ (without loss of generality, we suppose $i=1$; see Appendix A for more details)~\cite{LW20}. 
Particularly, when $I(\ketbra{\alpha}{\alpha},\{M_{\vec{x}}\},\{\vec{a}\})$ is very close to $C_q$, it turns out that $\epsilon_1$ and $\epsilon_2$ can be chosen such that $\epsilon_1/\epsilon_2\ll1$, implying that $\rho$ is close to a pure state~\cite{WL19}, which is consistent with the intuition mentioned above. 

With the estimations for $\hat{F}$ and $a_1$, if it holds that $\hat{F}\leq a_1$, a lower bound for the GME can be obtained as~\cite{LW20}
\begin{widetext}
 \begin{eqnarray*}
  E_{G}(\rho)\geq \max_{c\in\left[\frac{\hat{F}}{\sqrt{a_1}},\sqrt{a_1}\right]}\frac{a_1-c^2}{1-c^2}\left(1-\left(\frac{\hat{F}}{\sqrt{a_1}}c+\sqrt{1-\frac{\hat{F}^2}{a_1}}\sqrt{1-c^2}\right)^2\right).
\end{eqnarray*}
\end{widetext}

Actually, in addition to the GME, one can also lower bound the \emph{relative entropy of entanglement} (REE) $E_{R}(\rho)$ for $\rho$~\cite{VPRK97} by estimating $a_1$ and $\hat{F}$. Indeed, with the technique introduced in Ref.\cite{SSY+17}, the information on $a_1$ allows us to upper bound $S(\rho)$, the von Neumann entropy of $\rho$ that plays a key role in many-body systems ~\cite{NC00}. Combining this result with the information on $\hat{F}$, $E_{R}(\rho)$ can be directly lower bounded using the relation $E_{R}(\rho)\geq -2\log_2(\hat{F})-S(\rho)$ ~\cite{Wei08}.

Specifically, if $\rho$ is restricted to a $d\times d$-dimensional bipartite quantum state, 
the entanglement of formation (denoted as $E_f(\rho)$)~\cite{BDSW96} and the entanglement of distillation (denoted as $E_d(\rho)$) ~\cite{BDSW96} can also be quantified in a semi-DI manner. For this, first note that both of the two entanglement measures can be lower bounded by the \emph{coherent information} of $\rho$ defined as $I_C(\rho)=S(\rho_A)-S(\rho)$ ~\cite{SN96,Lloyd97}, i.e., $E_f(\rho)\geq E_d(\rho)\geq I_C(\rho)$ ~\cite{COF11}. Furthermore, the coherent information $I_C(\rho)$ can be lower bounded by upper bounding $S(\rho)$ and lower bounding $S(\rho_A)$ simultaneously from the correlation data $p(a_1a_2|x_1x_2)$ (the dimension $d$ of the bipartite system is known) ~\cite{WL19}. As a result, the entanglement of formation and distillation can be lower bounded semi-device-independently.

\begin{figure}[bhtp]
\begin{center}
\includegraphics [width= 1\columnwidth]{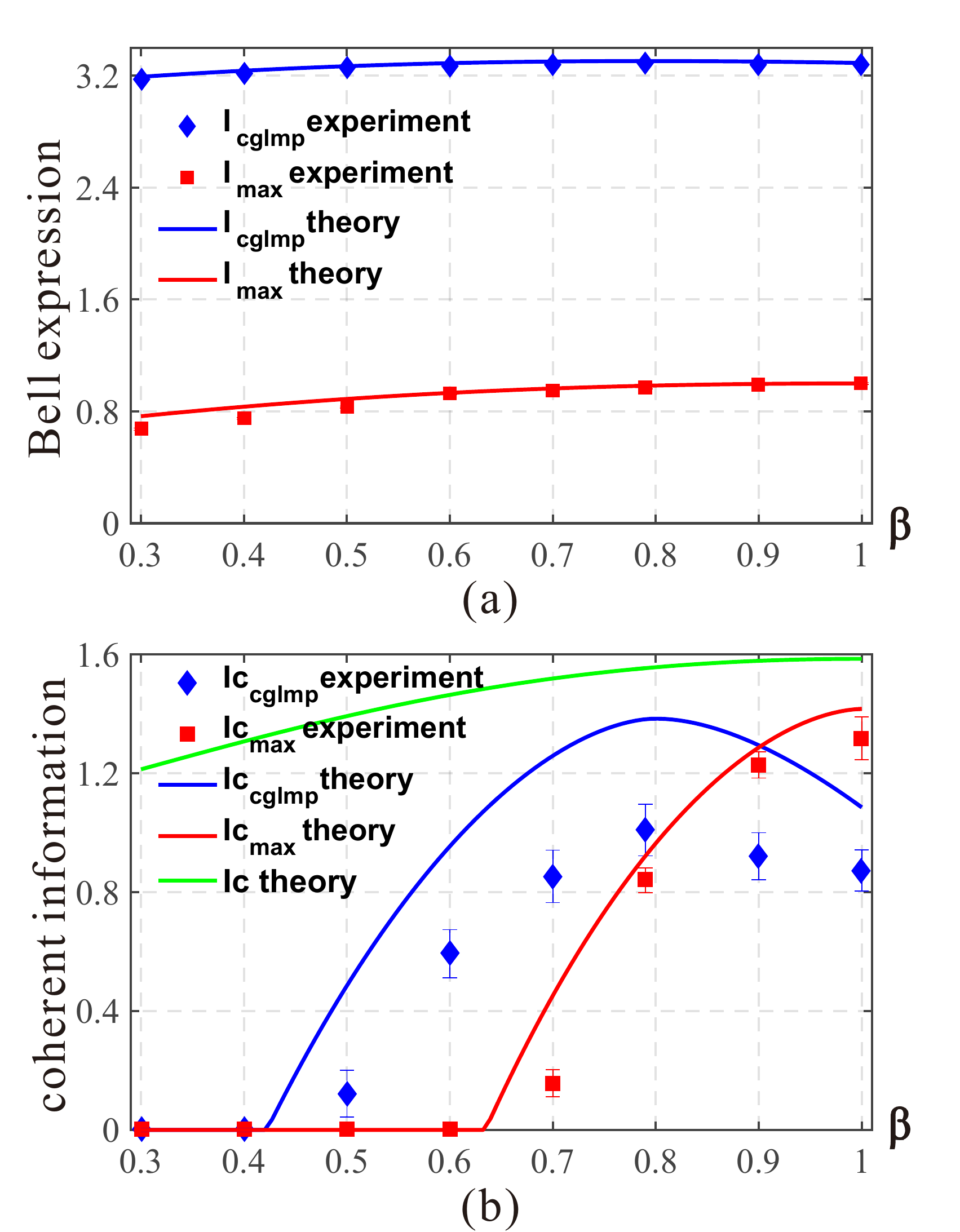}
\end{center}
\caption{Results of semi-DI entanglement quantification for qutrit-qutrit states, where $\beta\in[0.3,1]$. (a) Experimental observed Bell expressions of the CGLMP inequality and the inequality tailored for maximally entangled state are marked as blue and red dots respectively, matching well with the theoretical result (blue and red lines). (b) Coherent information as a lower bound of entanglement of the state $\ket{\Phi(\beta)}$. Experimental results are marked as blue and red dots for the two inequalities respectively, and the theoretical predictions using our method are plotted in the blue and red lines. For comparison, we also plot the exact coherent information of perfect $\ket{\Phi(\beta)}$ as the green line. The error bars in (a) are smaller than the marker size.
}
\label{fig:2}
\end{figure}

{\em Experimental implementation.---}The experimental setup to implement the trust-free entanglement quantification for multi-level and multi-partite quantum states is shown in Fig.~\ref{fig:1}. The setup mainly consists of entangled photon sources and measurement simulations for corresponding non-degenerate Bell-type inequalities.

In Fig.~\ref{fig:1}(a), we use a high-quality path-polarization hybrid encoded entanglement source~\cite{Hu2018beating} to generate desired entangled states beyond the qubit state space. In particular, two-qutrit states of the form $\ket{\Phi(\beta)}=(\ket{00}+\beta\ket{11}+\ket{22})/\sqrt{2+\beta^2}$ with varied $\beta$ are prepared by means of the process of degenerate spontaneous parametric down-conversion (SPDC). Here, the vertically-polarized (V) photon in the upper path is encoded as state $\ket{0}$, and the horizontally-polarized (H) and vertically-polarized photon in the lower path are encoded as state $\ket{1}$ and $\ket{2}$ respectively. The real coefficient $\beta$ is controlled by varying the angles of the half-wave plates (HWPs) at 404~nm. In Fig.~\ref{fig:1}(b), two ultra-bright beamlike EPR photon sources are used to generate the 3-partite Greenberger-Horne-Zeilinger (GHZ) state $\ket{\Psi}_3=(\ket{HHH}+i\ket{VVV})/\sqrt{2}$~\cite{chao16}. Here an HOM-interferometer ensures photons from different EPR sources are indistinguishable in arrival time, frequency and spatial degree of freedom, and the postselection on two events $\ket{HHHH}$ and $\ket{VVVV}$ results in a 4-photon GHZ state. The desired state $\ket{\Psi}_3$ can be obtained when one of the photons acts as a trigger and a phaser properly adjusts the relative phase between $\ket{HHH}$ and $\ket{VVV}$.

{\em Entanglement of qutrit-qutrit states.---}The previously introduced semi-DI entanglement quantification method is general and can be applied for any multi-level and multi-partite states. We first apply it on a $d\times d$ quantum system. Here both Alice and Bob are required to randomly perform two measurements on their qudits to test a Bell-type inequality. If we choose the inequality to be the 3-dimensional CGLMP inequality (or the Bell inequality tailored to maximally entangled states~\cite{SAT+17}), the involved four measurements have projection states admitting a general quantum-mechanical formula as
\begin{align*}
\ket{o(0)}&=\frac{1}{\sqrt{3}}(\ket{0}+e^{i\alpha1}\ket{1}+e^{i\alpha2}\ket{2}),\\
\ket{o(1)}&=\frac{1}{\sqrt{3}}(\ket{0}+e^{i(\alpha1+2\pi/3)}\ket{1}+e^{i(\alpha2+4\pi/3)}\ket{2}),\\
\ket{o(2)}&=\frac{1}{\sqrt{3}}(\ket{0}+e^{i(\alpha1+4\pi/3)}\ket{1}+e^{i(\alpha2+8\pi/3)}\ket{2}),
\end{align*}
where the phases $\alpha1,\alpha2\in[0, 2\pi)$. As depicted in Fig.~\ref{fig:1}(a), the above measurements can be realized via placing five phasers (P), five HWPs, two beam displacers (BDs), a polarizing beam splitter (PBS), and three single photon detectors sequentially. Specifically, the P2, P3 and P5 are set at $\alpha_1-\alpha_2$, $-\alpha_1$ and $-(\alpha_1+\pi/2)$, and the HWP1-5 are rotated at $45^{\circ}$, $67.5^{\circ}$, $72.37^{\circ}$, $45^{\circ}$ and $22.5^{\circ}$. The P1 and P4 set at 0 are used for temporal compensation and the detectors D1-D3 record three outcomes 0-2 respectively. Here the phaser consisting of two quarter-wave plates (QWPs) and an HWP can add an arbitrary phase between the H and V components.

As the first demonstration, we report our experimental results on the two bipartite Bell expressions in Fig.~\ref{fig:2}, where the values of Bell expressions can be seen in Fig.~\ref{fig:2}(a) and the lower bound for the coherent information can be seen in Fig.~\ref{fig:2}(b). Here the class of states we chosen is $\ket{\Phi(\beta)}$ with $\beta\in[0.3,1]$. As mentioned before, the coherent information is a lower bound for the entanglement of formation and the entanglement of distillation. In Fig.~\ref{fig:2}, our experimental data are marked with coloured points, while the theoretical predictions (produce quantum correlations using perfect quantum states and measurements, then apply our method if needed) are given as the coloured solid lines. Specifically, the blue points and line represent results for the CGLMP inequality, and the red points and line are for the inequality tailored for maximally entangled states. For comparison, we also plot the exact value of the coherent information as green solid line in Fig.~\ref{fig:2}(b) .
It can be seen that the measured Bell expressions match well with the theoretical lines, implying high-precision preparations and measurements of the qutrit-qutrit states. When choosing the CGLMP inequality, we obtain a maximal coherent information of $I_C=1.01\pm0.09$ for $\beta=0.79$, chiming with the trend of theoretical prediction. Additionally, the minimal $\beta$ in our experiment that we can set to certify entanglement is 0.5, while theoretically the coherent information should be positive when $\beta$ is larger than $\beta_{min}=0.4223$. See the Appendix for experimental results or more details on $a_1$ and $\hat{F}$.

A blemish of the CGLMP inequality when used as an entanglement quantifier is that the maximal violation is not obtained by the maximally entangled state. This can be avoided by utilizing the inequality tailored for maximally entangled states~\cite{SAT+17}. With this inequality, the detected coherent information increases with the parameter $\beta$ and a maximum of $I_C=1.32\pm0.07$ is obtained for maximally entangled qutrits, indicating a highly visible signal of entanglement beyond qubit systems. As a cost, the region of detectable states narrows down to about $\beta\geq0.64$, which is verified in our experiment, and we observe successfully the existence of entanglement at $\beta=0.7$.


\begin{figure}[bhtp]
\begin{center}
\includegraphics [width= 1\columnwidth]{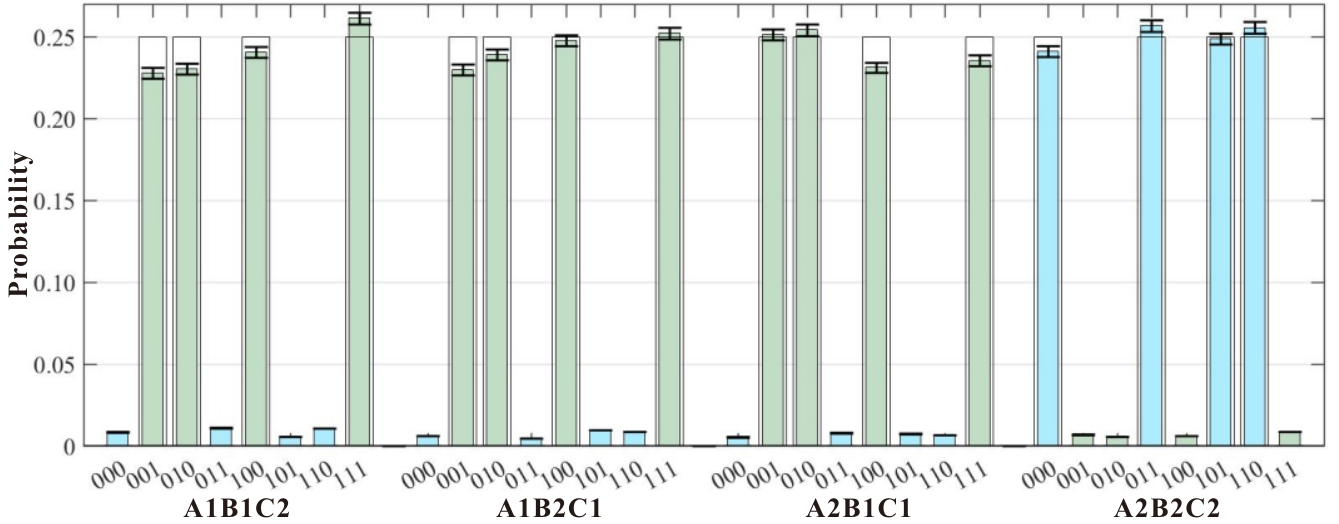}
\end{center}
\caption{Results on semi-DI entanglement quantification for the 3-qubit GHZ state. To test the 3-partite MABK inequality, Alice, Bob, and Charlie randomly perform Pauli X or Pauli Y measurements on their qubits. The coloured bars are the experimentally observed probabilities that obtain different outcomes on different measurement settings, with the corresponding theoretical predictions shown in gray edges. `A1B1C2' means Alice and Bob perform Pauli X and Charlie performs Pauli Y measurement, and `001' means their outcomes are -1, -1, 1 respectively. Light green and light blue bars represent that the number of outcome 1 is odd and even respectively. From these statistics, we obtain a value of the MABK expression $I_{MABK}=1.895\pm0.013$ and a corresponding lower bound for the GME $E_{G}(\ket{\Psi}_3)\geq0.169\pm0.006$.
}
\label{fig:3}
\end{figure}

{\em GME of the 3-qubit GHZ state.---}Then, we apply the method to quantify the entanglement of multi-partite system. We test the 3-partite MABK inequality on a 3-photon GHZ state $\ket{\Psi}_3$, where Alice, Bob, and Charlie randomly choose one of two Pauli measurements (Pauli-X and Pauli-Y) on their qubits. Single-qubit Pauli measurements can be achieved by an assemblage of a phaser, an HWP and a PBS. The measured statistics are recorded and later used to calculate the corresponding MABK expression, which allows us to lower bound the entanglement of the underlying quantum state.
As shown in Fig.~\ref{fig:3}, we list the measured statistics in coloured bars. From these correlations, we obtain an MABK inequality expression value of $I_{MABK}=1.895\pm0.013$ and a GME of $E_{G}(\ket{\Psi}_3)=0.169\pm0.006$, while the theoretical predictions are $2$ and $0.5$ respectively. Here, despite these mismatches, our results show enormous potential of non-degenerate Bell inequalities in quantifying multi-partite entanglement. The error bars of all the data are calculated from 100 simulations of Poisson statistics.

{\em Conclusion.---}{We have demonstrated semi-DI multi-level and multi-partite entanglement quantifications in a proof-of-principle experiment by preparing a class of entangled photonic qutrits and tripartite photonic GHZ states. Despite the detection loophole, our result, together with existing measurement-DI scenarios~\cite{LCQ12,BP12,Guo2019steering,Guo2020irreducible}, marks an important step towards complete DI entanglement quantification of quantum systems.}

\begin{acknowledgements}
This work was supported by the National Key Research and Development Program of China (No.\ 2017YFA0304100, No. 2016YFA0301300, No. 2016YFA0301700, and No. 2018YFA0306703), NSFC (Nos. 11774335, 11734015, 11874343, 11874345, 11821404, 11904357, and 20181311604), the Key Research Program of Frontier Sciences, CAS (No.\ QYZDY-SSW-SLH003), Science Foundation of the CAS (ZDRW-XH-2019-1), the Fundamental Research Funds for the Central Universities, Science and Technological Fund of Anhui Province for Outstanding Youth (2008085J02), and Anhui Initiative in Quantum Information Technologies (Nos.\ AHY020100, AHY060300).
\end{acknowledgements}

\appendix
\setcounter{figure}{0}
\renewcommand{\thefigure}{S\arabic{figure}}


\section{On the quantity $a_1$}
\begin{figure}[tbph]
\begin{center}
\includegraphics [width= 1\columnwidth]{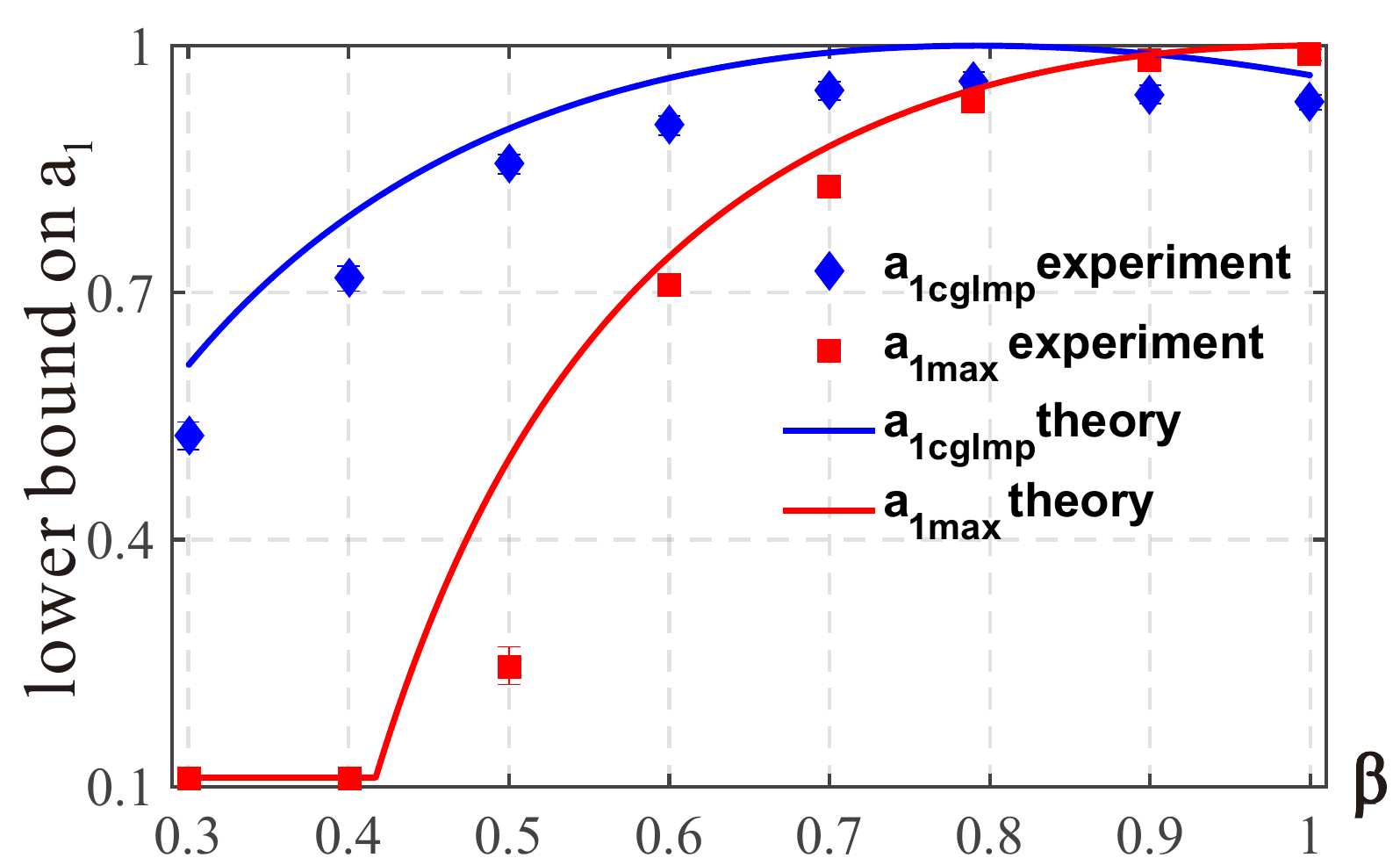}
\end{center}
\caption{Results of lower bounding $a_1$ for the qutrit-qutrit states $\ket{\Phi(\beta)}$ with $\beta\in[0.3,1]$. Experimental results are marked as blue and red dots for the two inequalities respectively, and the theoretical predictions (produce quantum correlations using perfect quantum states and measurements, then apply our method) are plotted in the blue and red lines. The error bars are calculated from 100 simulations of Poisson statistics.
}
\label{fig:4}
\end{figure}

As shown in the main text, the quantity $a_1$ is fundamental in quantifying the entanglement of unknown quantum states from probability distribution in semi-DI manner. In our experiment, the lower bounds for $a_1$ are obtained by applying the concept of non-degenerate Bell inequality directly. For example, the results of $a_1$ for the qutrit-qutrit demonstration can been seen in Fig.~\ref{fig:4}, where it can be seen that when the observed Bell value approaches the maximal, $a_1$ becomes closer and closer to 1.

\section{On the quantity $\hat{F}$}
Suppose that the probability distribution $p(\vec{a}|\vec{x})$ is obtained by measuring the target quantum state $\rho$ with local measurements $\{M_{\vec{x}}\}$. Let $\ket{\phi}$ be an $n$-partite pure product states, and $q^*(\vec{a}|\vec{x})$ be the correlation produced by measuring $\ket{\phi}$ with the same local measurements $\{M_{\vec{x}}\}$. Then there exist probability distributions $q^*_i(a_i|x_i)$ such that $q^*(\vec{a}|\vec{x})=\prod_{i=1}^{n}q^*_i(a_i|x_i)$, and for any $\vec{x}$ it holds that
\begin{align*}
	F(\ket{\phi}\bra{\phi},\rho)\leq F(q^*_{\vec{x}},p_{\vec{x}})= \sum_{\vec{a}}\sqrt{q^*(\vec{a}|\vec{x})p(\vec{a}|\vec{x})},
\end{align*}
where $p_{\vec{x}}\equiv p(\cdot|\vec{x})$, $q^*_{\vec{x}}\equiv q^*(\cdot|\vec{x})$, and the inequality comes from the fact that any quantum measurement cannot make the fidelity between two quantum states smaller. This means $F(\ket{\phi}\bra{\phi},\rho)\leq\min\limits_{\vec{x}}F(q^*_{\vec{x}},p_{\vec{x}})$, and
\begin{align*}
	F(\ket{\phi}\bra{\phi},\rho)\leq\max\limits_{q}\min\limits_{\vec{x}}F(q_{\vec{x}},p_{\vec{x}}),
\end{align*}
where the maximization is over product correlations $q$ and $q_{\vec{x}}\equiv q(\cdot|\vec{x})$. Combining this with the max-min inequality
\begin{align*}
	\max\limits_{q}\min\limits_{\vec{x}}F(q_{\vec{x}},p_{\vec{x}})\leq\min\limits_{\vec{x}}\max\limits_{q}F(q_{\vec{x}},p_{\vec{x}}),
\end{align*}
we have that
\begin{align*}
	F(\ket{\phi}\bra{\phi},\rho)\leq\min\limits_{\vec{x}}\max\limits_{q}F(q_{\vec{x}},p_{\vec{x}}).
\end{align*}

Therefore, we eventually get an upper bound for the fidelity between the target state and a pure product state, denoted as $\hat{F}$, based on the probability distribution $p(\vec{a}|\vec{x})$ only. Indeed, once $\vec{x}$ is fixed, the inner maximization can be computed using symmetric embedding~\cite{RV13} and the shifted higher-order power method (SHOPM) algorithm~\cite{KM11}, yielding a correct answer up to numerical precision with very high probability.



\begin{thebibliography}{9}

\bibitem{Gisin07}
N. Gisin and R. Thew, Nat. Photonics {\bf 1}, 165(2007).

\bibitem{BB84}
C. H. Bennett and G. Brassard, in Proceedings of the IEEE International Conference on Computers, Systems and Signal Processing, Bangalore, India, 1984 (IEEE, New York, 1984), pp. 175-179; IBM Tech. Discl. Bull. 28, 3153-3163 (1985).

\bibitem{Ekert92}
A. K. Ekert, Phys. Rev. Lett. {\bf 67}, 661 (1991).

\bibitem{Gisin02}
N. Gisin, G. Ribordy, W. Tittel, and H. Zbinden, Rev. Mod. Phys. {\bf 74}, 145 (2002).

\bibitem{oneway}
R. Raussendorf and H. J. Briegel, Phys. Rev. Lett. {\bf 86}, 5188 (2001).

\bibitem{Vidal03}
G. Vidal, Phys. Rev. Lett. {\bf 91}, 147902 (2003).

\bibitem{Srednicki94}
M. Srednicki, Phys. Rev. E {\bf 50}, 888 (1994).

\bibitem{rmp19}
D. A. Abanin, E. Altman, I. Bloch, and M. Serbyn, Rev. Mod. Phys. {\bf 91}, 021001 (2019).

\bibitem{bloch15}
 M. Schreiber, S. S. Hodgman, P. Bordia, H. P. L\"{u}schen, M. H. Fischer, R. Vosk, E. Altman, U. Schneider, and I. Bloch, Science {\bf 349}, 842 (2015).

\bibitem{xiaogang95}
 X.-G. Wen, Adv. Phys. {\bf 44}, 405 (1995).

\bibitem{xiaogang06}
M. Levin and X.-G. Wen, Phys. Rev. Lett. {\bf 96}, 110405 (2006).

\bibitem{kitaev06}
A. Kitaev and J. Preskill, Phys. Rev. Lett. {\bf 96}, 110404 (2006).

\bibitem{Guhne02}
O. G\"{u}hne, P. Hyllus, D. Bru{\ss}, A. Ekert, M. Lewenstein, C. Macchiavello, and A. Sanpera, Phys. Rev. A {\bf 66}, 062305 (2002).

\bibitem{Brunner08}
N. Brunner, S. Pironio, A. Ac\'{i}n, N. Gisin, A. A. M\'{e}thot, and V. Scarani, Phys. Rev. Lett. {\bf 100}, 230501 (2008).

\bibitem{Acin07}
A. Ac\'{i}n, N. Brunner, N. Gisin, S. Massar, S. Pironio, and V. Scarani, Phys. Rev. Lett. {\bf 98}, 230501 (2007).

\bibitem{Mayers04}
D. Mayers and A. Yao, Quantum Inf. Comput. {\bf 4}, 273 (2004).

\bibitem{Moroder13}
T. Moroder, J. D. Bancal, Y. C. Liang, M. Hofmann, and O. G\"{u}hne, Phys. Rev. Lett. {\bf 111}, 030501 (2013)

\bibitem{DIDW1}
J. Ahrens, P. Badziag, A. Cabello, and M. Bourennan, Nat. Phys. {\bf 8}, 592 (2012).

\bibitem{DIDW2}
M. Hendrych, R. Gallego, M. Mi\v{c}uda, N. Brunner, A. Ac\'{i}n and J. P. Torres, Nat. Phys. {\bf 8}, 588 (2012).

\bibitem{LHFBN}
B. Hensen et al., Nature {\bf 526}, 682 (2015).

\bibitem{DIRG}
Y. Liu et al., Nature {\bf 562}, 548 (2018).

\bibitem{Zhang18}
W.-H. Zhang, G. Chen, X.-X. Peng, X.-J. Ye, P. Yin, X.-Y. Xu, J.-S. Xu, C.-F. Li, and G.-C. Guo, Phys. Rev. Lett. {\bf 122}, 090402 (2019).

\bibitem{LCQ12}
H. K. Lo, M. Curty, and B. Qi, Phys. Rev. Lett. {\bf 108}, 130503 (2012).

\bibitem{BP12}
S. L. Braunstein and S. Pirandola, Phys. Rev. Lett. {\bf 108}, 130502 (2012).

\bibitem{MG12}
T. Moroder and O. Gittsovich, Phys. Rev. A {\bf 85}, 032301 (2012).

\bibitem{LVB11}
Y. C. Liang, T. V\'ertesi, and N. Brunner, Phys. Rev. A {\bf 83}, 022108 (2011).

\bibitem{CGP+02}
D. Collins, N. Gisin, S. Popescu, D. Roberts, and V. Scarani, Phys. Rev. Lett. {\bf 88}, 170405 (2002).

\bibitem{Mermin90}
N. D. Mermin, Phys. Rev. Lett. {\bf 65}, 1838 (1990).

\bibitem{Ardehali92}
M. Ardehali, Phys. Rev. A {\bf 46}, 5375 (1992).

\bibitem{BK93}
A. V. Belinski\u{i} and D. N. Klyshko, Phys. Usp. {\bf 36}, 653 (1993).

\bibitem{BH01}
D. C. Brody, L. P. Hughston, J. Geom. Phys. {\bf 38}, 19 (2001).

\bibitem{WG03}
T.-C. Wei, P. M. Goldbart, Phys. Rev. A {\bf 68}, 042307 (2003).


\bibitem{KM11}
T. G. Kolda and J. R. Mayo, SIAM J. Matrix Anal. Appl. {\bf 32}, 1095 (2011).

\bibitem{LW20}
L. Lin and Z. Wei, e-print arXiv:2008.12064.

\bibitem{WL19}
Z. Wei and L. Lin, e-print arXiv:1903.05303.

\bibitem{CHSH70}
J. F. Clauser, M. A. Horne, A. Shimony, and R. A. Holt, Phys. Rev. Lett. {\bf 24}, 549 (1970).

\bibitem{SG01}
V. Scarani and N. Gisin, J. Phys. A {\bf 34}, 6043 (2001).

\bibitem{VPRK97}
V. Vedral, M. Plenio, M. Rippin, and P. Knight, Phys. Rev. Lett. {\bf 78}, 2275 (1997).

\bibitem{SSY17}
G. Smith, J. A. Smolin, X. Yuan, Q. Zhao, D. Girolami, and X. Ma, e-print arXiv:1707.09928.

\bibitem{NC00}
 M. A. Nielsen, I. L. Chuang, Quantum Computation and Quantum Information, Cambridge University Press, 2000.

\bibitem{Wei08}
T.-C. Wei, Phys. Rev. A {\bf 78}, 012327 (2008).

\bibitem{BDSW96}
C. Bennett, D. DiVincenzo, J. Smolin, and W. Wootters, Phys. Rev. A {\bf 54}, 3824 (1996).

\bibitem{SN96}
B. Schumacher and M. A. Nielsen, Phys. Rev. A {\bf 54}, 2629 (1996).

\bibitem{Lloyd97}
S. Lloyd, Phys. Rev. A {\bf 55}, 1613 (1997).

\bibitem{COF11}
 M. F. Cornelio, M. C. de Oliveira, and F. F. Fanchini, Phys. Rev. Lett. {\bf 107}, 020502 (2011).

\bibitem{Hu2018beating}
X.-M. Hu, Y. Guo, B.-H. Liu, Y.-F. Huang, C.-F. Li, and G.-C. Guo, Sci. Adv. \textbf{4}, eaat9304 (2018).

\bibitem{chao16}
C. Zhang, Y.-F. Huang, C.-J. Zhang, J. Wang, B.-H. Liu, C.-F. Li, and G.-C. Guo,  Opt. Express {\bf 24}, 027059 (2016).

\bibitem{SAT+17}
A. Salavrakos, R. Augusiak, J. Tura, P. Wittek, A. Ac\'{i}n, and S. Pironio, Phys. Rev. Lett. {\bf 119}, 040402 (2017).

\bibitem{Guo2019steering}
Y. Guo, S. Cheng, X.-M. Hu, B.-H. Liu, E.-M. Huang, Y.-F. Huang, C.-F. Li, G.-C. Guo, and E. Cavalcanti, Phys. Rev. Lett. {\bf 123}, 17402 (2019).

\bibitem{Guo2020irreducible}
Y. Guo, B.-C. Yu, X.-M. Hu, B.-H. Liu, Y.-C. Wu, Y.-F. Huang, C.-F. Li, and G.-C. Guo, npj Quantum inf. {\bf 6}, 52 (2020).

\bibitem{RV13}
S. Ragnarsson and C. F. Van Loan, Linear Algebra Appl. {\bf 438}, 853 (2013).

\end{thebibliography}
\end{document}